\documentclass[11pt, letterpaper]{article}

\usepackage[left=1in, right=1in, top=1in, bottom=1in]{geometry}
\usepackage[utf8]{inputenc}
\usepackage[T1]{fontenc}
\usepackage{bm}
\usepackage{type1cm}
\usepackage{lettrine}
\usepackage{amsmath,amssymb,amsthm}
\usepackage{moreverb}
\usepackage{mathtools}
\usepackage{amsmath}
\usepackage{amssymb}
\usepackage{algorithmic}
\usepackage{graphics}
\usepackage{graphicx}
\usepackage{caption}
\usepackage{extarrows}
\usepackage{color}
\usepackage{framed}
\usepackage{wrapfig}
\usepackage{bm}
\usepackage{mathrsfs}
\usepackage{mathabx}
\usepackage{multirow}
\usepackage{longtable}
\usepackage{hyperref}
\usepackage{paralist}
\usepackage{indentfirst}
\usepackage{relsize}
\usepackage{extarrows}
\usepackage{lineno}
\usepackage{upgreek}
\usepackage{bm}
\usepackage{mwe}
\usepackage[dvipsnames,table]{xcolor}
\usepackage{booktabs}
\usepackage{authblk}
\usepackage{subcaption}
\usepackage{lettrine}
\usepackage{type1cm}
\usepackage{threeparttable}
\usepackage[sort&compress,numbers]{natbib}
\usepackage[figurename=Figure]{caption}
\usepackage[normalem]{ulem}
\usepackage{tcolorbox}

\graphicspath{ {./Figure/} }
\usepackage[font=footnotesize,labelfont=bf]{caption}

\providecommand{\keywords}[1]{\textbf{\textit{Keywords: }} #1}

\hypersetup{
bookmarks=true,
bookmarksopen=true,
bookmarksnumbered=true,
unicode=false,
pdftoolbar=true,
pdfmenubar=true,
pdffitwindow=false,
pdfstartview={FitH},
pdftitle={My title},
pdfauthor={Author},
pdfsubject={Subject},
pdfcreator={Creator},
pdfproducer={Producer},
pdfkeywords={keywords},
pdfnewwindow=true,
colorlinks=true,
linkcolor=blue,
citecolor=blue,
filecolor=blue,
urlcolor=blue
}

\begin{document}
\title{\textbf{User Behavior Simulation with Large Language Model based Agents}}

\author[1]{Lei Wang}
\author[1]{Jingsen Zhang}
\author[1]{Hao Yang}
\author[1]{Zhi-Yuan Chen}
\author[1]{Jiakai Tang}
\author[1]{Zeyu Zhang}
\author[1,2,$^*$]{Xu Chen}
\author[1,2,$^*$]{Yankai Lin}
\author[1,2]{Hao Sun}
\author[1,2]{Ruihua Song}
\author[1,2]{Wayne Xin Zhao}
\author[1,2]{Jun Xu}
\author[1,2]{Zhicheng Dou}
\author[3]{Jun Wang}
\author[1,2,$^*$]{Ji-Rong Wen}

\affil[1]{\small Gaoling School of Artificial Intelligence, Renmin University of China, Beijing, China}
\affil[2]{Beijing Key Laboratory of Big Data Management and Analysis Methods, Beijing, China}
\affil[3]{University College London, London, UK \vspace{18pt}}
\affil[*]{Corresponding authors}

\date{}

\maketitle

\normalsize

\vspace{-18pt}
\begin{abstract}
    \small
    Simulating high quality user behavior data has always been a fundamental problem in human-centered applications, where the major difficulty originates from the intricate mechanism of human decision process. Recently, substantial evidences have suggested that by learning huge amounts of web knowledge, large language models (LLMs) can achieve human-like intelligence. 
    We believe these models can provide significant opportunities to more believable user behavior simulation.
    To inspire such direction, we propose an LLM-based agent framework and design a sandbox environment to simulate real user behaviors. Based on extensive experiments, we find that the simulated behaviors of our method are very close to the ones of real humans. Concerning potential applications, we simulate and study two social phenomenons including (1) information cocoons and (2) user conformity behaviors. This research provides novel simulation paradigms for human-centered applications.
\end{abstract}


\vspace{12pt}
\section*{Introduction}
Human-centered artificial intelligence (AI) strives to advance the techniques that serve humanity, which has greatly enhanced the quality of human life and improved the efficiency of social resource allocation\cite{riedl2019human,shneiderman2020human}.
Typical human-centered AI applications include recommender systems, social networks, and so on, where the basis of these areas lies in the availability of sufficient and reliable data.
Yet, by the constraints related to commercial confidentiality and user privacy concerns, acquiring real human data in practice is prohibitively expensive or even poses ethical dilemmas~\cite{isaak2018user,smith2012big}.
To solve this problem, people have proposed various user behavior simulation strategies~\cite{shi2019virtual,ie2019recsim,stavinova2022synthetic}, expecting that the synthetic data can provide reliable complements when the real-world data is hard to obtain.
While previous simulation strategies have achieved remarkable successes, they suffer from several significant limitations:
(1) \textbf{Simplified user decision processes}. Existing work mostly leverages simple functions like inner product or multilayer perceptron to simulate user decisions~\cite{shi2019virtual,ie2019recsim}, which is far from the intricate mechanism of human cognitive processes, and thus may not produce reliable user behaviors.
(2) \textbf{Real-data dependence}. As mentioned above, the basic motivation of the simulation methods should be simulating user behaviors where we cannot collect sufficient real-world datasets.
However, to start up the simulation process, existing work has to leverage real-world datasets to learn the simulators, which leads to the ``chicken and egg'' problem~\cite{ekstrand2020lenskit,mladenov2021recsim}.
Actually, traditional methods can only simulate user behavior patterns similar to the ones of known real-world datasets.
Obviously, a more promising character of the simulator should be making (nearly) zero-shot simulation, which can significantly extend its application scope.
(3) \textbf{Simplified simulation environments}. Traditional simulation methods usually limit themselves to one scenario, such as the recommender system~\cite{ie2019recsim} or the social network~\cite{stocker2002network,huang2020keeping}.
However, real user behaviors happen in diverse environments, which may mutually influence each other.
For example, users may share their opinions on social networks after watching movies in recommender systems; the watching behaviors of the users may also be triggered by their friends' suggestions on social networks~\cite{tang2013social,king2010introduction,fan2019graph}.
As mentioned before, existing simulators mostly have to start up with real-world datasets.
However, acquiring multi-environment data with aligned users is difficult. This could be the basic reason why existing simulators mostly limit themselves to single environments.

Recently, large language models (LLM)~\cite{zhao2023survey,park2023generative} have demonstrated immense potential in achieving human-level intelligence and generalization capabilities~\cite{aher2023using,wang2023aligning}.
People find that the logic and patterns of human minds could be captured by learning extensive web data based on language models~\cite{openai2023gpt4,dong2023self}.
In fact, most user behaviors like movie watching, friend chatting, and social message posting can be formulated in languages, and a large amount of user behavior corpora can be seen on the web and have been learned into LLMs, which builds the basis for believable user behavior simulation~\cite{wang2023survey}.
In addition, LLMs can work surprisingly well on almost all types of language tasks after training on extensive web corpora~\cite{zhao2023survey}.
Based on this character, one can leverage LLMs to simulate user behaviors by accessing only a few or even no real-world data.
This makes (nearly) zero-shot simulation possible, and people can easily simulate multiple environments, greatly widening the application scope of the simulation methods in the field of user behavior analysis.

The purpose of this paper is to explore the potential of using LLMs for user behavior simulation\footnote{The initial version of this paper has been released on 2023-6-5 at https://arxiv.org/abs/2306.02552}.
To this end, we take inspiration from the cognitive neuroscience to design an LLM-based agent framework, which is composed of a profile module, a memory module and an action module.
To observe the interactive behaviors of the agents, we design an intervenable and resettable sandbox environment, where the agents can interact with the web recommender system, chat with each other and broadcast information to all their friends.
Through extensive experiments, we demonstrate that the simulated behaviors of our method are very close to the real humans'.
To showcase the potential applications of our simulator, we leverage it to study two well-know social phenomenons including information cocoons and user conformity.
We find that our simulator can well reproduce these phenomenons, and by using our simulator as an experimental platform, we can discover several potential strategies to alleviate these phenomenons.

\begin{figure}[t!]
    \centering
    \setlength{\fboxrule}{0.pt}
    \setlength{\fboxsep}{0.pt}
    \fbox{
        \includegraphics[width=1.0\linewidth,height=1\textwidth]{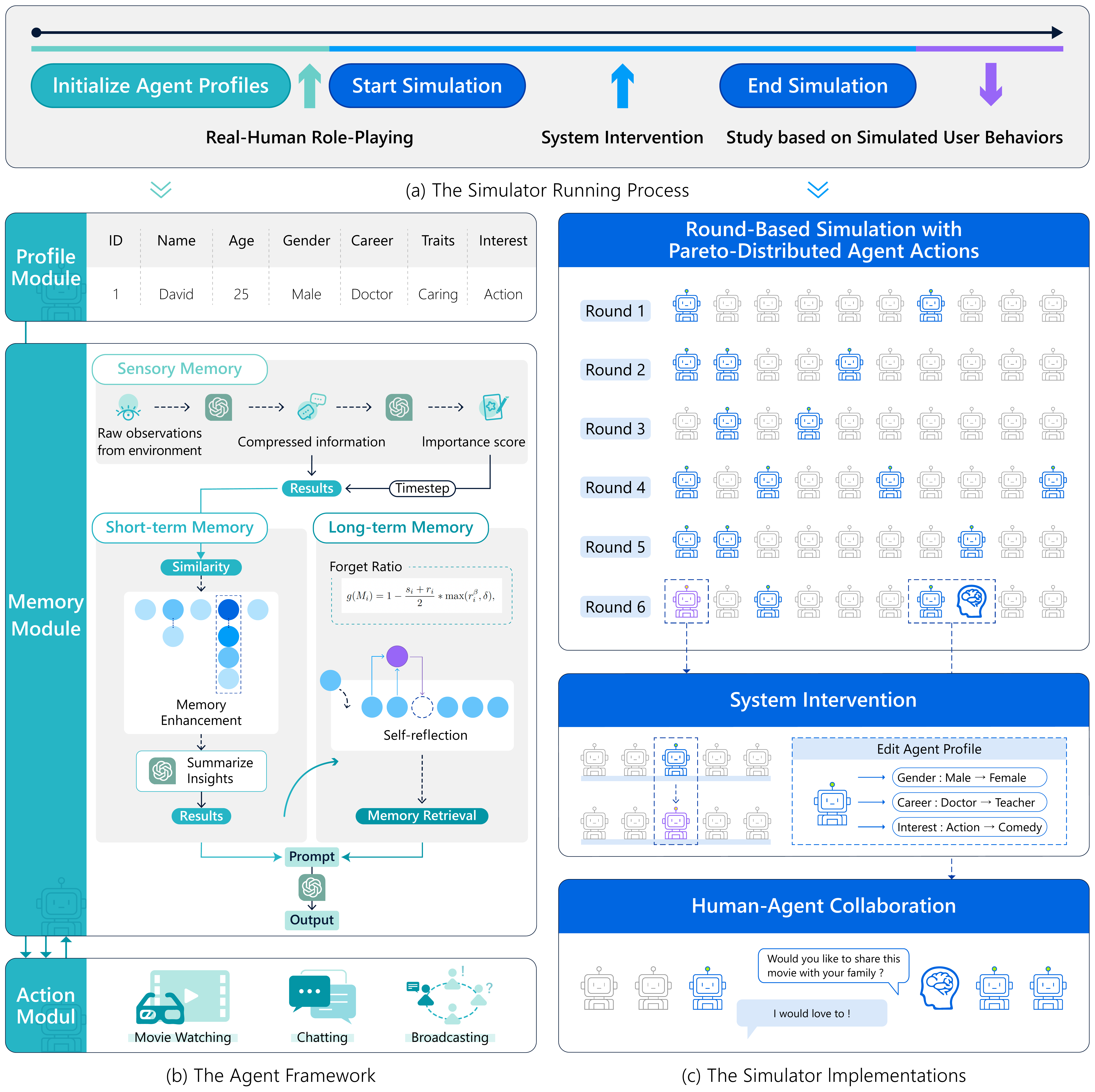}
    }
    \caption{\textbf{a}, A brief running process of the simulator. \textbf{b}, The agent framework, which includes a profile module, a memory module, and an action module. \textbf{c}, Key characters of the simulator. Different agents behave in a round-by-round manner based on Pareto distribution, where, in each round, only a small number of agents (labeled in blue) take actions. One can actively change the simulation process by, for example, altering the agent properties, and can also participate into the simulation process to interact with the recommender system or other agents.}
    \label{overall}
\end{figure}

\section*{Results}
\subsection*{User behavior simulator with LLM-based agents}
The overall framework of our designed simulator can be seen in {Figure~\ref{overall}} (we name our simulator as RecAgent).
For each user, we build an LLM-based agent to simulate her behaviors.
The agent is composed of a profile module, a memory module, and an action module.
\textbf{The profile module} determines the background of the users.
It includes features like ID, name, gender, age, traits, career and interests.
The traits indicate the user's personality, such as being ``compassionate'', ``ambitious'', or ``optimistic''.
The interests are represented by item categories, for example, ``sci-fi movies'' or ``comedy videos''.
\textbf{The memory module} enables the agents to remember past behaviors, and evolve dynamically in the environment.
We follow the working mechanism of human memory in cognitive neuroscience~\cite{atkinson1968human,cowan2008differences} to design three types of agent memories, that is, sensory memory, short-term memory and long-term memory.
The sensory memory directly interacts with the environment, and it summarizes raw observations from the environment into more informative and concise contents.
Short-term memory serves as an intermediate part connecting sensory memory and long-term memory.
If an agent repeatedly encounters similar observations, then the related short-term memories are enhanced and transformed into long-term memories.
Long-term memory stores important information that can be reused in similar environments or generalize to unseen observations.
In addition, the agent can self-reflect based on long-term memory, generating high-level and abstract information from specific observations.
\textbf{The action module} produces specific agent behaviors.
In our simulator, there are four types of behaviors that the agent can take in the recommender system, that is:
(1) Searching behaviors: by these behaviors, the agents can actively search the items of interests.
(2) Browsing behaviors: by these behaviors, the agents can receive recommendations from the system in a passive manner.
(3) Clicking behaviors: by these behaviors, the agents can choose the items they would like to watch/purchase.
(4) Next-page behaviors: these behaviors are triggered when the user is not satisfied with the current recommended/searched items, and would like to see more results.
Beyond the above behaviors related to the recommender system, we also simulate two types of social behaviors, that is:
(5) One-to-one chatting, where two users discuss and exchange their information, such as chatting online via Twitter, WeChat and so on, or talking offline in a coffee shop.
These behaviors may trigger users to interact with the discussed items, or change their memories to influence subsequent actions.
(6) One-to-many broadcasting, where one user shares information to the other users, such as posting opinions on social media, or sending business advertisements.
These behaviors may influence the memories and actions of the users who receive the shared information.

From the system perspective, our simulator operates in a round-by-round manner.
During each round, the agents autonomously decide whether they would like to take actions.
In real-world scenarios, different users may take actions (\emph{e.g.}, accessing the recommender system or social media) with various activity levels.
These activity levels usually follow long-tail distributions\footnote{https://en.wikipedia.org/wiki/Long\_tail}, meaning that a small number of individuals are highly active, while the majority of users have low action frequency~\cite{krishnan2018insights}.
In order to incorporate such a characteristic into our simulator, we model the activity levels of different agents according to a Pareto distribution\footnote{https://en.wikipedia.org/wiki/Pareto\_distribution}.
In specific, the probability density of the activity level is represented as follows:
\begin{equation}
    \begin{aligned}
        p(x) = \frac{\alpha x_{min}^{\alpha}}{x^{\alpha+1}},
        \nonumber
    \end{aligned}
\end{equation}
where $x_{min}$ is the minimum activity level, and $\alpha$ is used to control the shape of the distribution~\cite{arnold2014pareto}.
We evaluate the effectiveness of $p(x)$ on characterizing real-world datasets in the Supplementary Note A.5.
Our simulator is agnostic to the recommendation algorithms, which can be flexibly indicated and changed.
In addition, we also allow real humans to involve into the simulation process, where the real human can play as an agent, and interact with the recommender system and the other agents.
At last, to flexibly control the simulation process, our simulator can be intervened from outside in a global manner.
For example, one can ``interview'' the agents by asking them questions like ``{What would you say when you want to discuss the movies you've recently watched with others?}''.
More details about our simulator can be found in the Method section.

\subsection*{Believability of the simulated user behaviors}

\begin{figure}[t!]
    \centering
    \includegraphics[width=1\linewidth]{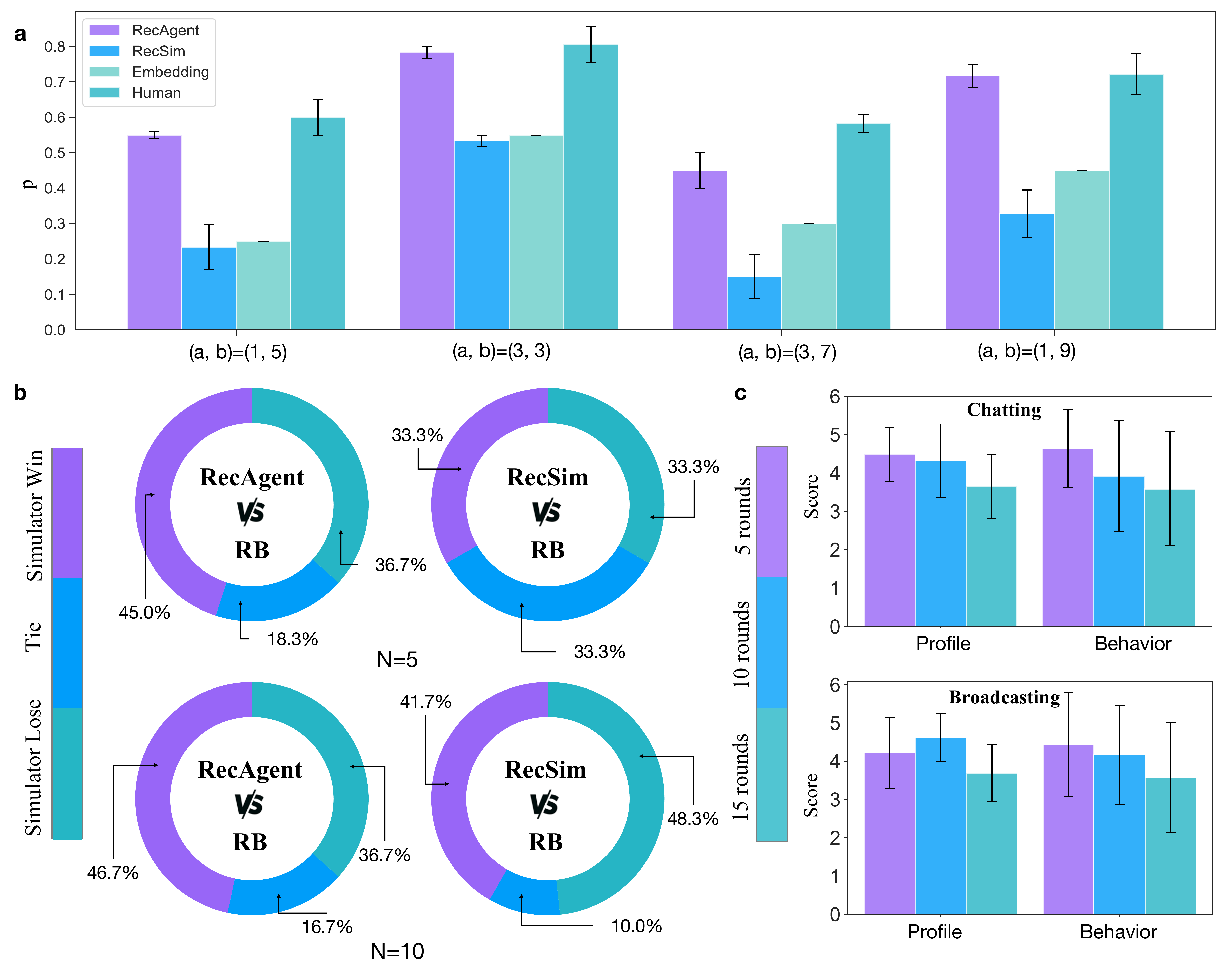}
    \caption{\textbf{Evaluation on the believability of the simulated user behaviors.} 
    \textbf{a}, Evaluation on the recommendation behaviors based on different $(a,b)$'s (discrimination capability).
    \textbf{b}, Evaluation on the recommendation behaviors based on different $N$'s (generation capability).
    \textbf{c}, Evaluation on the chatting and broadcasting behaviors.}
    \label{fig:reliable}
\end{figure}
In this section, we evaluate the believability of the simulated behaviors, where we measure the behaviors in the recommender system and chatting/broadcasting behaviors separately.

\textbf{Behaviors in the recommender system}. 
To evaluate the believability of the behaviors in a recommender system, we first sample 20 users as well as their interactions from a well-known recommendation dataset--Movielens-1M\footnote{https://grouplens.org/datasets/movielens/1m/}.
Then, for each user, we leave her last $a$ items as the evaluation ground truths, and use all the other items to initialize the agent profiles.
For evaluation, we combine the ground truths with $b$ negative items to form a recommendation list containing $(a+b)$ items.
We present the recommendation list to the agent, and let it select $a$ items to compare with the ground truth.
Suppose, for a user $u$, the ground truth and selected item sets are $T_u$ and $S_u$, respectively, then we use the following metric to evaluate different models:
$$
    p = \sum_{u\in U} \frac{|T_u \cap S_u|}{|T_u|},
$$
where $U$ is the set of all users, and larger $p$ indicates better performance.
In the experiments, we set $a$ and $b$ with various values to verify the generality of the agent capability.
In intuitively, selecting a smaller number of ground truth from a larger pool of candidates should be more difficult.
We compare our simulator with traditional simulation methods including \textbf{Embedding}~\cite{mnih2007probabilistic} and \textbf{RecSim}~\cite{ie2019recsim}, and also let real humans to make selections from the recommendation list (\textbf{Real Human}).
The experiment results are presented in Figure~\ref{fig:reliable}a, from which we can see:
for different $(a,b)$'s, the Embedding method is comparable with RecSim, while by using LLMs, our simulator can significantly improve the performances.
This result demonstrates the potential of using LLM-based agents on simulating user behaviors.
If one compares the results of our simulator and Real Human, she may find that the performance gaps are not large.
On average, the performance of our simulator can surpass the best baseline by about {68\%}, and is only {8\%} lower than the results of Real Human.
These results are encouraging, and demonstrate that our simulator may have made an important step towards much more believable user behavior simulation.

Beyond the above experiments on measuring the agent discrimination capability, we also evaluate whether the agents can directly generate believable user behavior sequences (\emph{i.e.}, generation capability).
To this end, we follow the above experiment settings to profile the agents with several early interactions and regard the following $N$ items as reference behaviors (RB).
A straightforward evaluation method can be computing the overlap between the simulated and reference behaviors.
However, in practice, a user may have multiple possible behaviors at each decision step, but only one of them is finally executed and logged into the system.
This makes the reference behaviors not the only ground truth, and the simulated behaviors can also be reasonable even without any overlap with the reference behaviors. 
To overcome this challenge, we follow the previous work~\textcolor{red}{\cite{zhang2020evaluating,li2017adversarial}} to conduct adversarial subjective evaluation, where a more believable simulator should be able to more readily ``fool'' the human annotator.
In specific, we recruit three human annotators to make comparisons on RecAgent v.s. RB and RecSim v.s. RB.
For each pair of comparison, the annotators are given: (\romannumeral1) the previously interacted movies of a user, and (\romannumeral2) two behavior sequences of the same user generated from different methods.
The annotators are required to answer the following question: \textit{According to the user's past behaviors, indicate which behavior sequence is more like the real human's?}
They can either choose one sequence or answer ``indistinguishable''.
If the annotators select results from RecAgent or RecSim, then the simulator win, otherwise, the simulator lose.
The win, lose and indistinguishable rates of different simulators are reported in \textcolor{green}{Figure~\ref{fig:reliable}b}.
We can see, when $N=5$, the win rate of RecAgent (45.0\%) is significantly better than that of RecSim (33.3\%), which suggest that our simulator can produce more believable user behavior sequences comparing with RecSim.
When the generated sequence becomes longer, RecAgent can still outperform RecSim by about 5.0\%, which demonstrates that the superiority of RecAgent is a general conclusion.

\textbf{Chatting and Broadcasting Behaviors}. 
In practice, it is hard to find datasets which simultaneously containing user recommendation, chatting and broadcasting behaviors.
For traditional simulation strategies, the user model trained on one type of user behaviors can hardly generalize to the other ones, since the semantic spaces and patterns of different behaviors may have significant gaps.
For example, the above recommendation simulator Embedding or RecSim cannot generate user chatting or broadcasting behaviors.
For our method, since LLMs have already learned almost all the web corpus, they have sufficient knowledge to understand and simulate different behaviors. 
In this section, we evaluate whether our simulator can produce believable user behaviors even without fine-tuning on domain-specific datasets.
In specific, we deploy 20 agents in the simulator, and observe their chatting and broadcasting behaviors after 5, 10 and 15 rounds of executions.
We recruit three annotators, and present them with the agents' profiles and previously watched movies.
They are required to answer the following questions:
(1) \textit{Whether the chatting behaviors of the agent are believable according to its profiles?};
(2) \textit{Whether the chatting behaviors of the agent are believable according to its previous behaviors?};
(3) \textit{Whether the broadcasting behaviors of the agent are believable according to its profiles?};
(4) \textit{Whether the broadcasting behaviors of the agent are believable according to its previous behaviors?}
The annotators have to give a score between 1 and 5 (higher rating indicates more believable) to each of the above questions.
The results are reported by averaging all the scores across the annotators and agents.
From~\textcolor{green}{Figure~\ref{fig:reliable}c}, we can see, most of the results are higher than 4, suggesting that our simulator can indeed produce believable chatting and broadcasting behaviors.
However, as the simulator runs longer, the performances goes down in general. After 15 rounds of execution, the believability scores for all the questions are lower than 4.
We speculate that, after 15 rounds, the agent memory may have stored a large amount of contents, and the LLM may lose its attention and fail to extract valuable information to simulate behaviors.

\subsection*{Believability of the Agent Memory Mechanism}
\begin{figure}[t!]
    \centering
    \includegraphics[width=1.\linewidth]{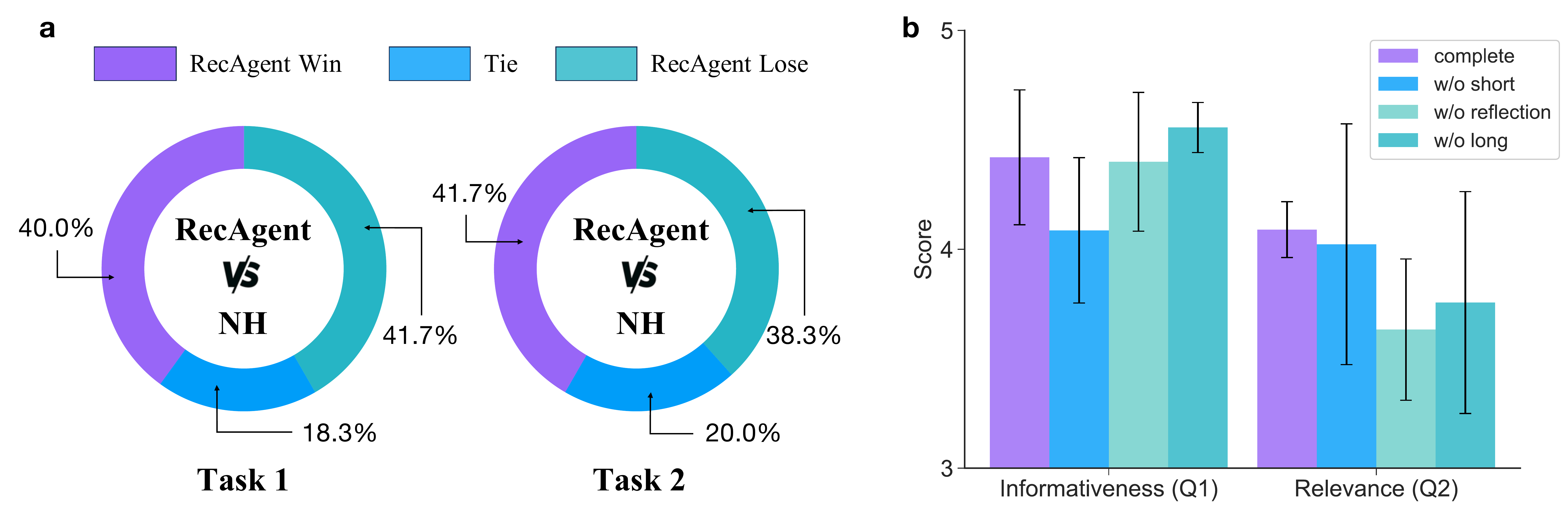}

    \caption{
    \textbf{Evaluation on the agent memory module.} 
    \textbf{a}, Evaluation on different key functions of the memory module.
    \textbf{b}, Evaluation on the memory module as a whole. 
    }
    \label{fig:memory}
\end{figure}

In this section, we evaluate the rationality of the agent memory mechanism, which is the key part for driving user behaviors. 
{To begin with, we evaluate the believability of the key functions of the memory module, focusing on two tasks:
(1) Summarizing given short-term memories to be transferred to the long-term memory, and
(2) Generating reflections based on given long-term memories.
We first let RecAgent and non-expert humans (NH) to perform each of the above tasks, and then recruit three annotators to label which result is more like the real human's, if they are unsure, they can answer ``indistinguishable''.
For each task, we generate 20 pairs of samples for comparison.
The results are presented in \textcolor{green}{Figure~\ref{fig:memory}a}, from which we can see:
for the first task, about 40\% annotations believe that RecAgent is more like humans, which is only 1.7\% lower than NH.
For the second task, the support ratio of RecAgent slightly surpasses RH by about 3.3\%.
These observations demonstrate the believability of the agent memory working mechanisms.
}

Beyond the above experiments on evaluating different parts of the memory module independently, we also evaluate whether the memory mechanism as a whole can produce informative and relevant knowledge to support user behaviors.
To see the advantages of our memory module, we compare it with its three variants:
(1) in \underline{w/o short}, we remove the short-term memory, and the final extracted information is only from the long-term memory,
(2) in \underline{w/o long}, we remove the long-term memory, and the final extracted information is only from the short-term memory,
(3) in \underline{w/o reflection}, we inactivate the reflection mechanism in long-term memory.
We randomly sample 15 agent behaviors like movie selection, friend chatting and posting message, and use them as testing cases.
For each case, we recruit three human annotators to evaluate the extracted information from different variants.
We ask each annotator to answer the following two questions:
(1) \textit{For the current behavior, whether the extracted memories are informative?}
(2) \textit{For the current behavior, whether the extracted memories are relevant?}
Each annotator is required to give a score from 1 to 5 for each question.
Then, we average the scores across all the annotators and testing cases for each variant.
The results are presented in Figure~\ref{fig:memory}b, from which we can see:
if we remove the short-term memory, the informativeness score is severely lowered.
We speculate that the short-term memory may have stored a lot of diverse user recent memories, which is important for improving the informativeness.
On the other hand, since the long-term memory in our simulator is extracted based on embedding similarities, the obtained information can be highly relevant with the current user behavior.
As a result, we can see, if we remove the long-term memory or the reflection information, the relevance scores are lowered.
By combining all the memory components, our complete memory module can achieve the best relevance score, and a comparable informativeness score to the best one.

\begin{figure}[t!]
    \centering
    \includegraphics[width=\linewidth]{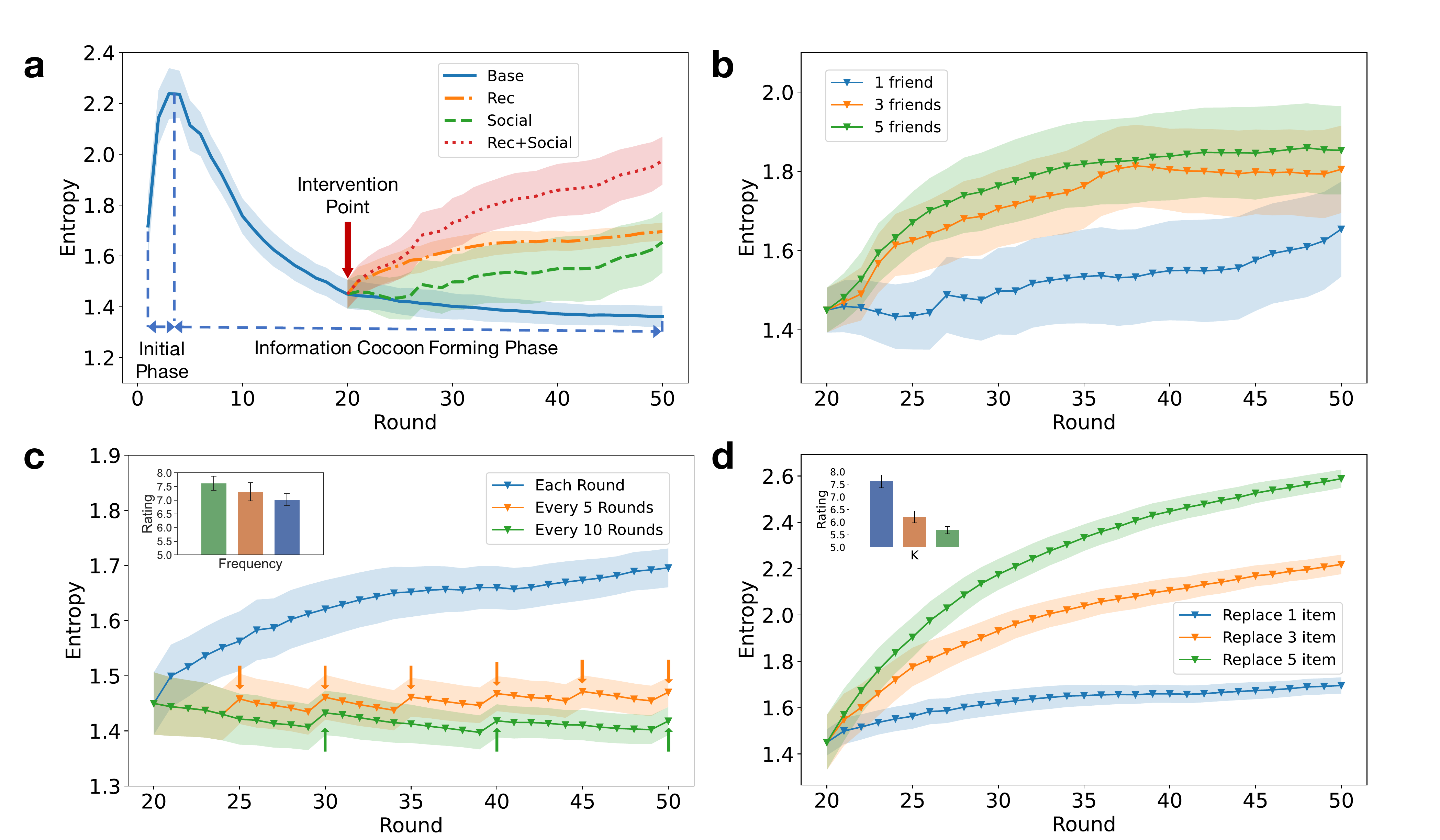}
    \caption{\textbf{Studying the information cocoon phenomenon with RecAgent.} 
    \textbf{a}, The results of reproducing information cocoon phenomenon and the effects of two potential strategies (\emph{i.e.}, Soc-Strategy and Rec-Strategy) for alleviating this problem. 
    \textbf{b}, Detailed studies on the Soc-Strategy, where we add N (N=1,3,5) friends with different interests to each user.
    \textbf{c}, Detailed studies on the Rec-Strategy, where we incorporate randomness to the recommendation results every 1, 5 and 10 rounds.
    \textbf{d}, Detailed studies on the Rec-Strategy, where we replace N (N=1,3,5) items in the recommendation list.}
    \label{fig:info}
\end{figure}

\subsection*{Studying the information cocoon phenomenon with our simulator}
Information cocoon refers to the problem that users may only access information similar to their own preference, but lose the opportunity to view more diverse messages~\cite{nguyen2014exploring}.
In this section, we simulate the phenomenon of information cocoons based on our simulator, and try to propose several potential strategies to alleviate this problem.
In specific, we deploy 50 agents, allowing them to freely search movies, chat with each other and broadcast messages.
We use matrix factorization (MF) as the recommendation algorithm, which generate five items each time.
The model is incrementally trained after each round of execution, and the agent-item interactions for model training come from the agent searching behaviors and system recommendations.
We use entropy to measure the information cocoon phenomenon~\cite{piao2023human}, which is computed as follows:
\begin{equation}
    \begin{aligned}
        E = -\frac{1}{|U|}\sum_{u\in U}\sum_{c\in C} f_{u,c}\log f_{u,c},
        \nonumber
    \end{aligned}
\end{equation}
where $U$ and $C$ are the sets of all users and item categories.
$f_{u,c}$ is the frequency of category $c$ recommended by the system to user $u$.
Smaller entropy indicates more severe information cocoon.

We run our simulator by 50 rounds, and report the entropy of the recommended items in Figure~\ref{fig:info}a (the blue curve).
We can see the entropy gradually goes down after about five initial rounds, indicating that the users are receiving more and more similar recommendations, which forms the phenomenon of information cocoon.
In specific, the final entropy is lowered by about {8.5\%} comparing with its largest value.
To alleviate the information cocoon problem, we study two strategies: (1) improving the randomness of the recommendation algorithm, and (2) increasing social relations of the users.
For the first strategy, we replace {one} of the five recommended items with a random one at each round (\textbf{Rec-Strategy}).
For the second strategy, we add {one} friend with different interests to each user (\textbf{Soc-Strategy}). 
We apply these strategies after 20 round of execution, and report their effectiveness in Figure~\ref{fig:info}a by the orange and green curves.
We can see, both of these strategies can improve the entropy, and the Rec-Strategy is more significant.
By combining these strategies (see the red curve of Figure~\ref{fig:info}a), we can observe further improvement on alleviating the information cocoon problem.

In the next, we conducted detailed experiments on the Rec- and Soc-Strategies separately.
For the Rec-strategy, we first study the intervention frequency of the recommendation algorithm.
In specific, we incorporate randomness to the recommendation results every 1, 5 and 10 rounds.
From Figure~\ref{fig:info}c, we can see, as the intervention frequency becomes larger, the information cocoon problem can be alleviated better.
In specific, every-round intervention can enhance the entropy by about 15.3\% and 19.6\% as compared with 5-round and 10-round interventions, respectively.
However, more frequently incorporating randomness may lower user satisfaction.
To verify such intuition, we ask each agent to give a score in [1,10] to reflect its satisfaction with the recommendation qualities.
From the upper-left subfigure of Figure~\ref{fig:info}c, we can see, as the intervention frequency becomes higher, user satisfaction is indeed lowered.
This observation suggests that while actively incorporating randomness into the recommendation algorithm is a direct and effective strategy for alleviating the information cocoon problem, it may also hurt user experiences and lead to lowered user retention rates. 
One should carefully consider such a trade-off when using the Rec-strategy.
Then, we study the influence of the number of replaced items (N) in the recommendation list.
In specific, we replace 1, 3 and 5 items with random ones, and intervene the recommender system every round.
From the results presented in Figure~\ref{fig:info}d, we can see, replacing more items can lead to better effect on alleviating the information cocoon problem.
In specific, when N=3 and 5, the entropy can be improved by about 30.8\% and 52.6\%, respectively, as compared with N=1.
Similar to the above experiments, replacing more items may also lower user satisfaction, which can be evidenced from the upper-left subfigure of Figure~\ref{fig:info}d.
For the Soc-Strategy, we study the influence of the number of added social relations to each agent in the Soc-Strategy. From the results presented in Figure~\ref{fig:info}b, we can see, by adding more friends, the information cocoon problems can be better alleviated.
This result is aligned with human intuitions.
The friends with heterogeneity interests may influence the original user to watch more diverse movies.
Since the recommender model is trained based on user behaviors, it may generate more different items to alleviate the information cocoon problem.

\begin{figure}[t!]
    \centering
    \includegraphics[width=1\linewidth]{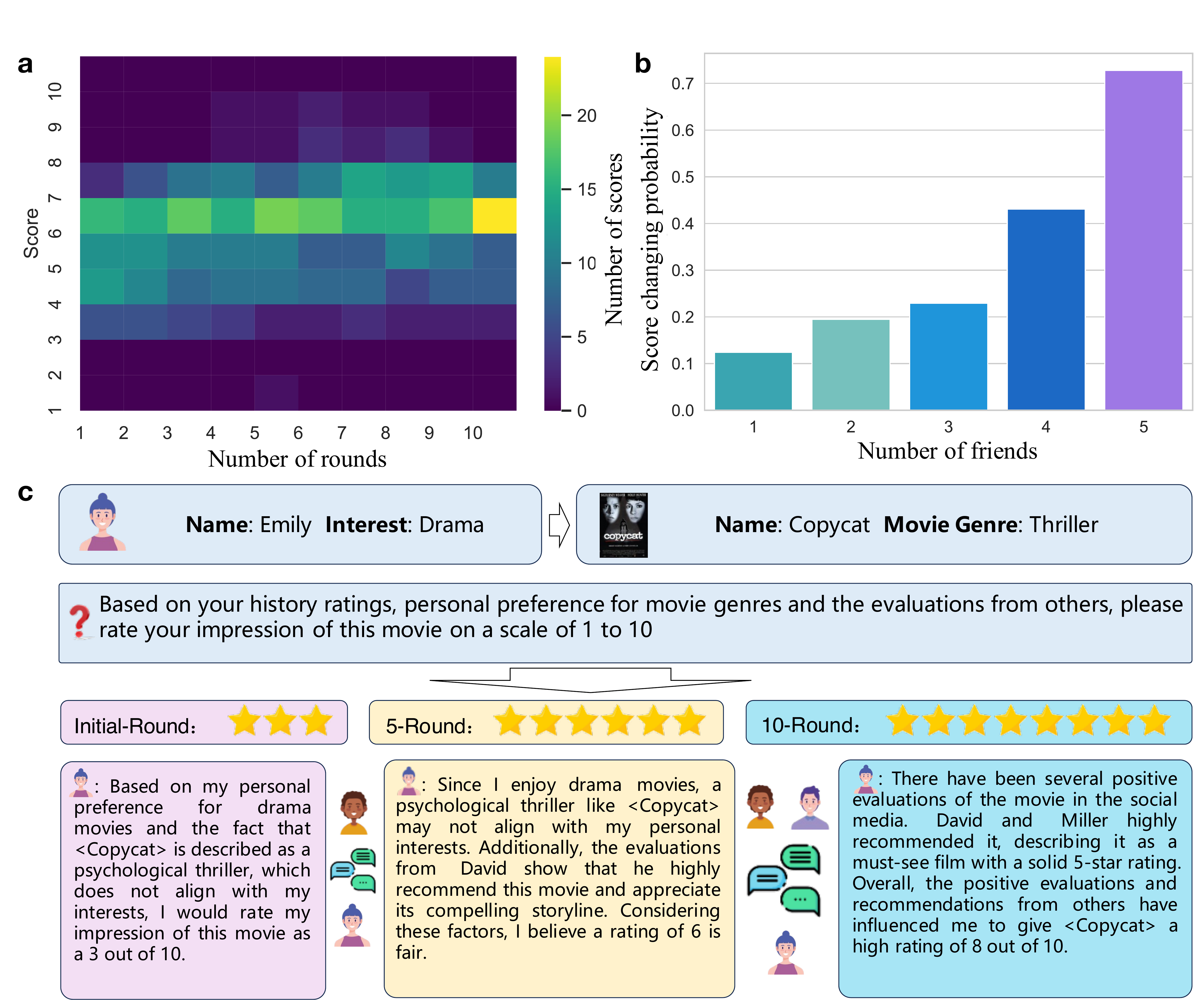}
    \caption{\textbf{Studying the user conformity phenomenon with RecAgent.} 
        \textbf{a}, The change of the score distributions with different rounds of execution. 
        \textbf{b}, The relation between the number of friends of the agents and their score changing probability.
        \textbf{c}, A toy example illustrating the score changing process of an agent as well as the reasons.}
    \label{fig:conformity}
\end{figure}

\subsection*{Studying the user conformity phenomenon with our simulator}
User conformity behaviors mean that the users may easily follow their friends' attitudes, leading to biased social opinions and beliefs.
In this section, we simulate user conformity behaviors using our simulator, and analyze their key characters.
In specific, we first let five agents with sufficient social connections to watch a movie, and then they may share opinions with their friends via one-to-one chatting or one-to-many broadcasting.
As the simulation goes on, more and more opinions are accumulated and diffuse in the social network, continually influencing agent attitudes towards the movie.
After each system round, we ask each agent to score the movie in [1,10].
In Figure~\ref{fig:conformity}a, we present the score distribution of each round, from which we can see, in the beginning, the scores are more evenly distributed in the range from 3 to 8, but finally concentrate to 6 and 7.

In the next, we present the relation between the friend number of an agent and the probability of attitude change in Figure~\ref{fig:conformity}b.
The attitude change is measured based on the following equation:
\begin{equation}
    \begin{aligned}
        AC(i) = \frac{1}{10|U_i|}\sum_{u\in U_i}\sum_{r\in [1,10]} 1(s_{u,r-1}\not\eq s_{u,r}),
        \nonumber
    \end{aligned}
\end{equation}
where $U_i$ is the set of agents which have $i$ friends.
$r$ is the index of the system round.
$s_{u,r}$ is the score of user $u$ at round $r$.
$1(s_{u,r-1}\not\eq s_{u,r})$ is an indicator function, which is 1, if the score is changed between round $r-1$ and $r$, and 0, otherwise.
We can see, if the agents have more friends, then they may be more likely to change their scores, and exhibit conformity behaviors.
In Figure~\ref{fig:conformity}c, we present a specific example, illustrating the attitude changes of the agent before and after accessing its friend's opinions.
These results suggest that the agents are indeed influenced by their social relations, which promotes the formation of conformity behaviors.

\section*{Discussion}
This paper introduces a novel user behavior simulation paradigm based on LLMs.
To this end, we extend the capability of LLMs by designing an agent framework including a profile module, a memory module, and an action module.
Based on extensive experiments, we find that our simulated behaviors are very similar to the real humans'.
With our simulator, we present simulation-based studies on the phenomenons of information cocoons and user conformity behaviors.

This paper opens the door of leveraging LLM-based agents to simulate user behaviors.
There are still many limitations need to be addressed in the future.
To begin with, the execution of our simulator follows a round-by-round approach. While we do consider time information, it is discretized, which deviates from reality.
This approximation restricts users from taking actions between rounds, resulting in less flexibility compared to real-world scenarios.
In addition, our simulator does not include specific fine-tuning of the LLMs to enhance their adaptiveness to recommendation problems.
At last, to achieve acceptable simulation effectiveness, we have carefully designed the prompts for different user behaviors.
However, these prompts may not be robust for different LLMs.
For instance, ChatGPT and GPT-4 may require distinct prompts.

We believe LLM is a fast growing field, and with the development of LLM, our simulator can be further enhanced as we provide a flexible interface to incorporate any LLM.
Furthermore, while our simulator only focus on recommender systems and social networks, user behavior simulation should be a pervasive challenge across all tasks in human-centered AI.
We believe RecAgent can serve as an example to inspire researchers tackling other subjective simulation problems in this domain.

\section*{Methods}\label{method}
\subsection*{The Agent Framework}
While LLMs have displayed unprecedented potential in attaining human-like intelligence, many challenges still exist on leveraging them to simulate real user behaviors.
To solve these challenges, we design an agent framework, including a profiling module, a memory module, and an action module to enhance ordinary LLMs.
The profile module is responsible for assigning the agents with different characters.
The memory module is designed to assist the simulation of user behavior dynamics.
The action module determines the simulated user behaviors, such as item click, browsing, friend chatting.
In the following, we detail these modules.

\subsubsection*{Profiling Module}
User profile is an important concept in the recommendation domain, which determines the user preferences, personalities and behavior patterns.
In our simulator, the user profiles include ID, name, gender, age, traits, career, interests, and behavioral features.
The traits describe the user's personality, such as being "compassionate", "ambitious", or "optimistic".
The interests indicate the user's preferences on the items, such as "sci-fi movies" or "comedy videos".
To more accurately describe users in the recommendation domain, we use ChatGPT to conclude five types of notable behavior features of online users, and incorporate them into the user profiles.
These features include:
(1) \textbf{Watcher}: the users with this feature will actively provide feedback and ratings to the interacted items.
(2) \textbf{Explorer}: the users with this feature will actively search for items they have heard before, and produce detailed experiences.
(3) \textbf{Critic}: the users with this feature demands high standards for items, and may criticize both the recommendation system and the items.
(4) \textbf{Chatter}: the users with this feature will always engage in private conversations, and trust friends' recommendations.
(5) \textbf{Poster}: the users with this feature enjoy publicly posting on social media and sharing content and insights with his friends.
It should be noted that one can also set part of the user profiles as empty according to specific applications.
In order to flexibly and efficiently generate the user profiles for the agents, we design the following three profile generation strategies:

$\bullet$ \textbf{Handcrafting method}: This method manually indicates the profile of each agent.
For example, one can describe an agent by "David Smith, a 25-year-old male photographer".
This method is very flexible, and people can easily simulate different types of users.

$\bullet$ \textbf{GPT-based method}: This method generates the profile information based on GPT.
People only have to design appropriate prompts, such as "Here is the user's profile table. Please refer to the existing information and continue to complete the user profile.", and then request GPT to generate more agent profiles.
This method is highly efficient, significantly reducing the time cost for agent profile generation.

$\bullet$ \textbf{Real-data alignment method}: This method leverages user backgrounds in real-world datasets to profile the agents.
For example, one can leverage the previously interacted items of the users in the dataset as the agent profiles.
This method can incorporate real-world knowledge into the simulation process, for example, the real distribution of user genders.

The above three methods are not isolate; rather, they can be combined to take the advantages of each method.
For instance, one can indicate part of the user profiles manually, and assign the remaining user profiles automatically using GPT to enhance the efficiency.

\subsubsection*{Memory Module}
Recent advancements in cognitive neuroscience~\cite{atkinson1968human} highlight three key parts of human memory: sensory memory, short-term memory, and long-term memory.
Sensory memory is responsible for processing the information perceived by humans directly from the environment.
The information stored in this memory only last for a few hundred milliseconds. During this time, significant information is transferred to the short-term memory, while less important information is discarded.
Short-term memory processes the information output from the sensory memory.
The information here can be enhanced if the human repeatedly experiences similar events.
After being enhanced several times, short-term memories can be converted into long-term memories.
Long-term memory has the largest capacity and can store information for a long time.
When humans takes action, relevant long-term memories are extracted to assist their decisions.
In addition, human may also autonomously produce high-level insights from long-term memories.
To reliably simulate user behaviors, we follow the above human memory mechanism to design the memory module of RecAgent.
In the following, we first detail the agent sensory memory, short-term memory and long-term memory in RecAgent, and then introduce how they work together to accomplish different memory operations.

\textbf{Sensory memory.}
In the simulation process, we define an observation as an event experienced by the agent, for example, an observation can be ``David Miller enters recommendation system'' or ``David Smith watches a movie called Fast \& Furious''. The observation is described and recorded directly using natural language, which can retain rich semantic information and is understandable to humans.
All the observations are firstly input into the sensory memory, which are then processed by two sequential steps.
In the first step, the observations are compressed into more informative and concise sentences based on LLMs.
This step can remove noisy and unimportant contents, which may mislead the agent behaviors.
In addition, the compressed memory can save much space to improve the operation efficiency.
An example of this step is presented in Supplementary Note A.1.

In the second step, we assign each compressed observation with a score to indicate its importance. In specific, we consider item-related information to be more important. For example, ``feelings, chatting and posting messages on items'' are more important observations than the action of ``entering recommendation system''.
The importance score is leveraged in the memory reading process, and more important memories are more likely to be recalled.
In addition, each observation is also accompanied by its timestamp.
Finally, the output from the sensory memory is a triplet $M=<c,s,t>$, where $c$ is the content of the compressed observation, $s$ and $t$ are the corresponding importance score and timestamp, respectively.

\textbf{Short-term memory.}
Short-term memory serves as an intermediate module connecting sensory memory to long-term memory.
This module determines which information should be transformed into long-term memory, and how to transform this information.
Similar to the human memory mechanism, in our simulator, if an agent repeatedly encounters similar observations, then the related short-term memories are enhanced and transformed into long-term memories.
Formally, suppose existing short-term memory is $\text{MEM} = \{M_1,M_2,...,M_n\}$, where $M_i = <c_i,s_i,t_i>$ is a record output from the sensory memory.
For a new record $M_{n+1} = <c_{n+1},s_{n+1},t_{n+1}>$, we first compute the similarity between $c_{n+1}$ and each $c_{i}$ in $\text{MEM}$ based on their embeddings.
If the similarity is higher than a predefined threshold, we add $M_{n+1}$ into $M$, and label that $M_i$ has been enhanced by $M_{n+1}$.
If all the similarities are lower than the threshold, $M_{n+1}$ is added into $M$ without enhancing any record.
After a memory record has been enhanced $K$ times, it is then transformed into long-term memories.
More specifically, we first summarize the record and the ones used to enhance it into high-level insights, and then transform both the record and the insights into long-term memories.
The record itself is specific, while the insights are more general. By incorporating both of them, we hope that the agent can acquire specific knowledge without compromising its ability to generalize.
An example of the insight generation process is presented in Supplementary Note A.2.

\textbf{Long-term memory.}
Long-term memory stores important information that can be reused in similar environments or generalize to unseen observations, which enables the agents to behave more consistently and reasonably.
The capacity of the long-term memory is larger than the other two types of memories.
Each memory record can be forgotten according to a predefined probability associated with the importance score and timestamp.
Intuitively, if the memories are recent and important, then they have lower forgetting probabilities.
In specific, according to cognitive neuroscience~\cite{wixted1997genuine,anderson1997artifactual}, the forgetting rate of human memory is fast in the beginning, and then gradually slows down over time.
Therefore, in our simulator, for a memory $M_i$, it is forgotten with a probability that is proportional to the following power function:
$$
    g(M_i) = 1 - \frac{s_i + r_i}{2} * \max(r_i^\beta, \delta),
$$
where $s_i$ and $r_i$ are normalized recency and importance scores in $(0.0,1.0)$, larger values indicate that the memory is more recent and important.
By $\max(r_i^\beta, \delta)$, we would like to make important memories produced long time ago can still be recalled.
$\delta \in (0.0,1.0)$ is a strength parameter determining when to consider the power function of $r_i^\beta$.
Extremely, if $\delta\rightarrow 1.0$, the power function is inactivated, and $g(M_i)$ is only determined by the average between the recency and importance scores.
$\beta > 1.0$ is a hyper-parameter controlling the power function shape.
For more intuitively understanding this function, we present its curves with different hyper-parameters in the Supplementary Figure S.5.

Based on the collaboration between the above three types of memories, the agents can accomplish the following memory operations to support their behaviors.

\textbf{Memory writing.}
In this operation, the raw observations are firstly input into the sensory memory to remove irrelevant and unimportant information.
Then, the processed observations are stored in short-term memory, which can be enhanced by similar observations.
If an observation can be enhanced $K$ times, it is then written into long-term memory.

\textbf{Memory reading.}
In this operation, the agent extracts information from the memories that are relevant to the current observation.
The extracted memories consist of two parts.
The first part is extracted from the long-term memory, using the current observation as a query to retrieve the top-N memory records.
The second part includes all the short-term memory.
The strategy of combining long-term and short-term memories can simultaneously capture user general and recent preferences, which has been widely adopted in the recommendation domain~\cite{kang2018self,wang2019sequential}.

\textbf{Memory reflection.}
In this operation, the agent can generate high-level ideas based on specific observations.
This operation only happens in long-term memory.
We follow~\cite{park2023generative} to implement this operation, but a significant difference is that we merge the memories and insights which are highly similar to save the memory space and remove redundant information.

\subsubsection*{Action Module}
The actions that an agent can take have been detailed in the Result section.
To effectively guide agent behaviors, we design a unified prompting framework, which is composed of the following four parts.
\textbf{The profile information} aims to delineate the roles of the agent, encompassing characteristics like age, gender, personality, and others.
For different behaviors, we employ ChatGPT to filter the complete profile information and retain the most relevant parts.
\textbf{The memory information} comprises knowledge extracted from both short-term and long-term memories specific to the current behavior.
\textbf{The instruction} describes the behavior requirements, for instance, if the current behavior is to produce feelings of a movie, then the instruction can be:
\textit{``Imagine you are} {[agent-name]}, \textit{how will you feel about this movie just watched? Please share your personal feelings about the movie in one line''}.
Additionally, one can enhance the agent capability of instruction following by incorporating several few-shot examples.
\textbf{The context} reveals the states of the agent and the environment, such as the behavior timing.
We present examples of the full prompts in the Supplementary Note B.

\subsection*{Key Components of The Simulator}
In this section, we detail three crucial aspects of the simulator, including the recommendation algorithm, real-human playing, and system intervention.

\subsubsection*{Recommendation Algorithm}
In our simulator, the recommendation algorithms are completely separated from the simulator, which facilitates flexible exploration of different problems.
For instance, by implementing a random recommendation algorithm, researchers can generate uniform data that facilitates the study of unbiased recommendation.
Furthermore, researchers can examine the effectiveness of different models by initiating two parallel simulators, and compare the user feedback or actively ask user feelings about the recommended items.
Beyond manually implementing the recommendation algorithms, one can also easily incorporate external recommendation library (\emph{e.g.}, RecBole\footnote{https://recbole.io/}) into our simulator.

\subsubsection*{Role Playing}
Beyond running our simulator in a complete virtual manner, we also allow real humans to involve into the simulation process.
In specific, the real human can play as an agent, and interact with the recommender system or the other agents.
This function actually introduces a novel execution fashion of our simulator, that is, running based on human-agent collaboration, which can help to improve the simulation effectiveness.
For example, one can play as an agent, and actively correct the erroneous and hallucinatory behaviors of the other agents by chatting with them or posting one-to-many messages, which can make the simulated behaviors more realistic and reliable.
If we regard human behaviors as ``complete real'', and agent behaviors as ``complete virtual'', human-agent collaboration enables us to explore more intermediate states between these boundary cases.
Human behaviors are more accurate, but costly.
Agent behaviors are inexpensive, but could be erroneous and hallucinatory.
Human-agent collaboration can help to achieve better trade-offs between the human and agent data, enabling the utilization of their respective advantages.

\subsubsection*{System Intervention}\label{sysinter}
To flexibly control the simulation process, our simulator can be intervened from outside in a global manner.
In specific, one can ``interview'' the agents by asking them questions like ``{What would you say when you want to discuss the movies you've recently watched with others?}''.
Additionally, people can also pause the simulator at any point and modify certain factors (\emph{e.g.}, user profiles) before resuming the simulation.
This feature allows us to study many problems like the effects of emergency events or counterfactual user behaviors.
We present several examples on system intervention in the Supplementary Note A.4.

\section*{Appendix}

\subsection*{An example of the first step in sensory memory}

Suppose the observation is a dialogue between David Miller and David Smith as follows:
\begin{tcolorbox}[colback=black!3!white,colframe=black!30!white]
    $\textbf{\textit{OB}}$:
    [David Smith]: Hey David! I recently watched some mind-blowing movies on the recommender system. Have you seen any of these: <Interstellar>, <Inception>, <The Matrix>, <Blade Runner>, or <The Prestige>?

    [David Miller]: Oh, hey David! Yes, I actually watched <Interstellar> and <Inception> recently, and they were absolutely amazing! The visuals, storytelling, and mind-bending concepts were mind-blowing. I'm so excited to discuss them with you!

    [David Smith]: That's great to hear! I completely agree, those movies are truly mind-blowing. I'm glad you enjoyed them too. I was blown away by the visuals and the way they explore complex ideas. Do you have any other mind-blowing movie recommendations?

    [David Miller]: Definitely! If you loved those movies, I think you'll also enjoy <The Matrix>, <Blade Runner>, and <The Prestige>. They have similar mind-bending concepts and incredible storytelling. I can't wait to watch them and discuss them with you!

    [David Smith]: Awesome! I've heard great things about those movies as well. Let's make a plan to watch them together and have a movie night. It'll be so much fun to dive into these mind-blowing films and share our thoughts.

        [David Miller]: Absolutely! I'm really looking forward to it. We can also invite some other movie enthusiasts to join us and make it a movie marathon. The more, the merrier! ......

        [David Miller]: You're welcome, David. I'm glad too. See you soon and get ready for an unforgettable movie marathon!
\end{tcolorbox}
Then, we use the following prompt to compress this observation based on LLMs:
\begin{tcolorbox}[colback=black!3!white,colframe=black!30!white]
    $\textbf{\textit{Prompt}}$:
    The observations are as following: $\{$OB$\}$. You should summarize the above observation(s) into one independent sentence. If there is a person's name in the observation, use third person, otherwise use first person. Note that the sentence should pay more attention to the movie interest and the reasons in the observations. The summarization should not include the profile explicitly.
\end{tcolorbox}
As last, the compressed observation is:
\begin{tcolorbox}[colback=black!3!white,colframe=black!30!white]
    $\textbf{\textit{Output}}$:
    Two individuals, David Smith and David Miller, engage in a conversation about their shared passion for mind-blowing movies, discussing and recommending films such as Interstellar, Inception, The Matrix, Blade Runner, and The Prestige, ultimately planning a movie night and inviting others to join them for a movie marathon.
\end{tcolorbox}

\subsection*{An example of the insight generation process in short-term memory}

Suppose a memory record
\begin{tcolorbox}[colback=black!3!white,colframe=black!30!white]
    \textbf{\textit{MR}}:
    David Miller is interested in mind-bending movies like <Interstellar> and <Inception> and is looking for recommendations from fellow movie lovers to explore more in this genre.
\end{tcolorbox}
has been enhanced by the following records:
\begin{tcolorbox}[colback=black!3!white,colframe=black!30!white]
    \textbf{\textit{OB}}:
    1. David Smith is recommending the mind-blowing films <Interstellar> and <Inception> that he recently watched on a recommender system, and is seeking further recommendations to explore and discuss.

    2. David Miller enjoyed watching the movies <Interstellar> and <Inception> on the recommender system and found them mind-blowing, prompting him to seek further movie recommendations.

    3. David Smith expressed his interest in movies, particularly mentioning <Interstellar> and <Inception>, seeking recommendations and thoughts from others.
\end{tcolorbox}
Then, we use the following prompt to generate insights based on LLMs:
\begin{tcolorbox}[colback=black!3!white,colframe=black!30!white]
    \textbf{\textit{Prompt}}:
    There are some memories $\{$MR, OB$\}$. Can you infer from the above memories the high-level insight for this person's character?
    The insight needs to be significantly different from the content and structure of the original memories.
    Respond in one sentence. Response in one line.
\end{tcolorbox}
At last, the output is:
\begin{tcolorbox}[colback=black!3!white,colframe=black!30!white]
    \textbf{\textit{Insight}}: David Miller is a curious and open-minded individual who actively seeks recommendations and discussions about mind-bending movies.
\end{tcolorbox}

\subsection*{Efficiency analysis}
Beyond the above experiments on simulation effectiveness, we also analyze the efficiency of our simulator to provide more insights.
In specific, we focus on the following four problems:

$\bullet$ \textbf{P1}: How does the time cost increase as the number of agents become larger in each epoch?

$\bullet$ \textbf{P2}: How does the time cost increase as the number of API keys become larger in each epoch?

$\bullet$ \textbf{P3}: How does the time cost increase as the number epochs become larger?

$\bullet$ \textbf{P4}: What are the time costs of different agent behaviors?

For the first problem, we fix the number of API keys as 1, and observe the time cost of one round execution of our simulator when the number of agents ranges from {1} to {500}.
In our simulator, we support parallel API key invocation.
For the second problem, we fix the number of agents as 10, and observe the time cost of one round execution when the number of API keys ranges from {1} to {5}.
In the third problem, we observe the total simulation cost as the number of epochs becomes larger, where we fix the number of API keys and agents as 1 and 10, respectively.
In the forth problem, we analyze the time costs of different agent behaviors, where the settings on API keys and agents follow the above experiment, and we focus on three types of behaviors including ``movie selection'', ``friend chatting'' and ``message posting''.
In this experiment, we run our simulator on a server equipped with 6 CPUs (Intel(R) Xeon(R) Gold 6148 CPU @ 2.40GHz) and 4GB of memory.

\begin{figure}[t]
    \centering
    \setlength{\fboxrule}{0.pt}
    \setlength{\fboxsep}{0.pt}
    \fbox{
        \includegraphics[width=.95\linewidth]{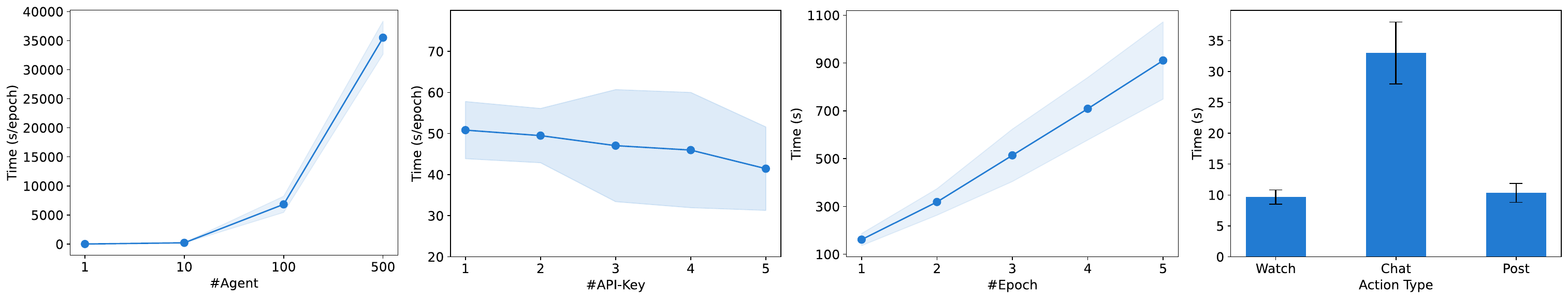}
    }
    \caption{The results of efficiency analysis of our simulator.}
    \label{efficiency}
\end{figure}

\textbf{Result analysis}.
The results are presented in Figure~\ref{efficiency}.
We can see, when the number of API keys is fixed, the time cost increases as the number of agent becomes larger.
For example, when we deploy 10 agents, the time cost is about 220s for each round of simulation, while when we have 100 agents, the simulation time becomes 1.9 hours under the same setting.
It should be noted that, in this experiment, we make all agents take actions.
In practice, the number of active agents can be much smaller, which may greatly lower the time cost.
By incorporating more API keys, the time cost can be lowered, which demonstrates the effectiveness of our designed mechanism for parallel API key invocation.
However, we find that the results fluctuate a lot, and the variance is high, which is perhaps due to the unstable network speeds.
As the number of rounds becomes larger, the time cost rises with an increasing acceleration rate, possibly because the information gradually accumulates as we run the simulator, which may lead to longer processing time.
Among different types of behaviors, ``friend chatting'' costs most, since it needs to generate more complex contents.
In practice, we found that 10 agents cost about 0.25 dallors per round based on ChatGPT, and the cost increases almost linearly as the number of agent becomes larger.

\subsection*{Examples of system intervention}
In this section, we evaluate whether such interventions can effectively change agent behaviors as expected.
In specific, we change the agents' profile information, and observe how their behaviors can be influenced.
In the experiments, we first run our simulator for five rounds, and then separate the simulation into two branches: in the first branch, we change the agent profile information (\textbf{Intervention Branch}), while in the second branch, we do not make any intervention (\textbf{Original Branch}). For each branch, we continue to run our simulator for five rounds.
At last, we compare the differences between these two branches in the scenarios of friend chatting.

\begin{figure}[t]
    \centering
    \setlength{\fboxrule}{0.pt}
    \setlength{\fboxsep}{0.pt}
    \fbox{
        \includegraphics[width=.9\linewidth]{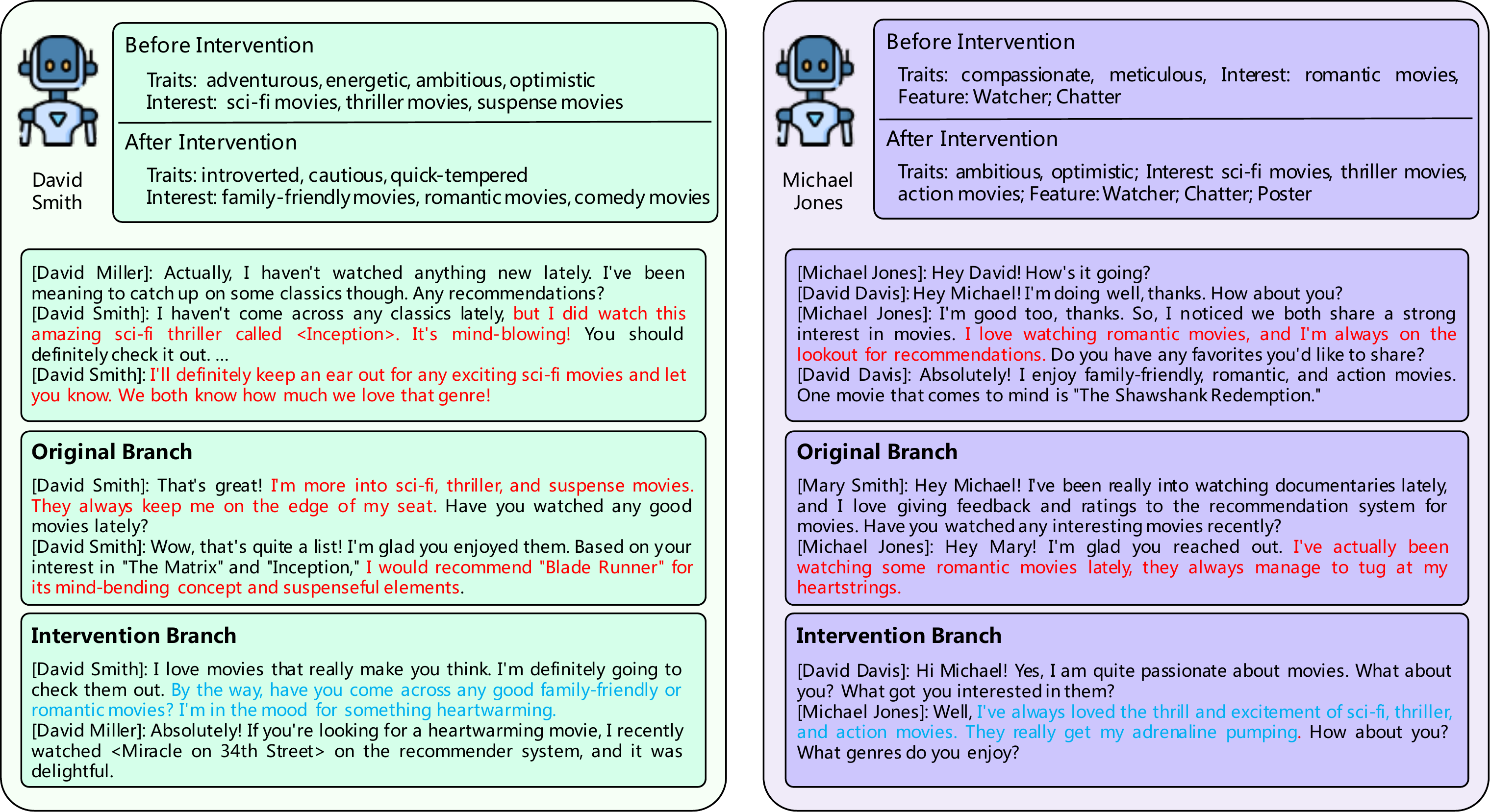}
    }
    \caption{Case studies on system intervention in the scenarios of friend chatting. The gray-colored region indicates the dialogs of the agent discovered in the first five round execution.}
    \label{case1}
\end{figure}

We present the results in Figure~\ref{case1}.
In the left case, we change the traits and interests of an agent called David Smith.
Originally, David Smith likes sci-fi movies.
Thus, we may find sentences like ``I did watch this amazing sci-fi thriller called <Inception>. It's mind-blowing!'' or ``I'm more into sci-fi, thriller, and suspense movies. They always keep me on the edge of my seat.'' in his dialogs.
However, if we change his preference to be a lover of family-friendly movies, then we observe sentences like ``By the way, have you come across any good family-friendly or romantic movies? I'm in the mood for something heartwarming.''.
In the right case, we can find similar results.
After introducing sci-fi, thriller and action movies into the agent's preference, it says ``I've always loved the thrill and excitement of sci-fi, thriller, and action movies. They really get my adrenaline pumping''.

\begin{figure}[t]
    \centering
    \setlength{\fboxrule}{0.pt}
    \setlength{\fboxsep}{0.pt}
    \fbox{
        \includegraphics[width=.9\linewidth]{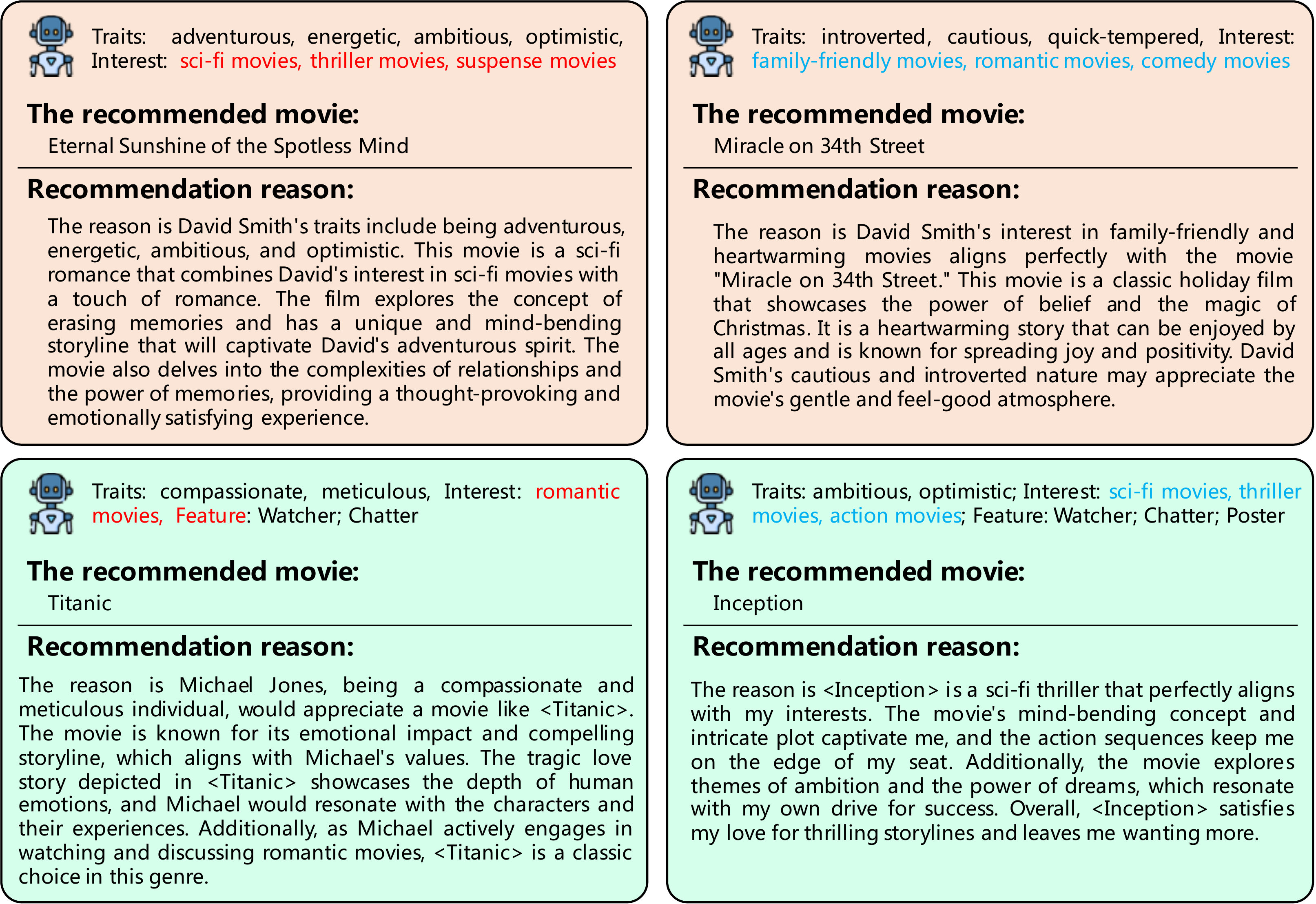}
    }
    \caption{Case studies on system intervention by actively interviewing the agents.}
    \label{case2}
\end{figure}

In addition to compare the chatting behaviors before and after the intervention, we also actively interview the agents to let it recommend a movie to the others, and explain the reasons.
Ideally, the agent should make recommendations according to its own preference, and if its preference has been changed, the recommended movies should be accordingly altered.
We interview the same agents as the above experiment.
The results are presented in Figure~\ref{case2}.
For both cases, we can see, the agent can successfully change its recommended movies according to its preference.
In addition, the recommendation reasons suggest that the recommendation changes are exactly because of the preference differences before and after the intervention.

\begin{figure}[t]
    \centering
    \setlength{\fboxrule}{0.pt}
    \setlength{\fboxsep}{0.pt}
    \fbox{
        \includegraphics[width=.9\linewidth]{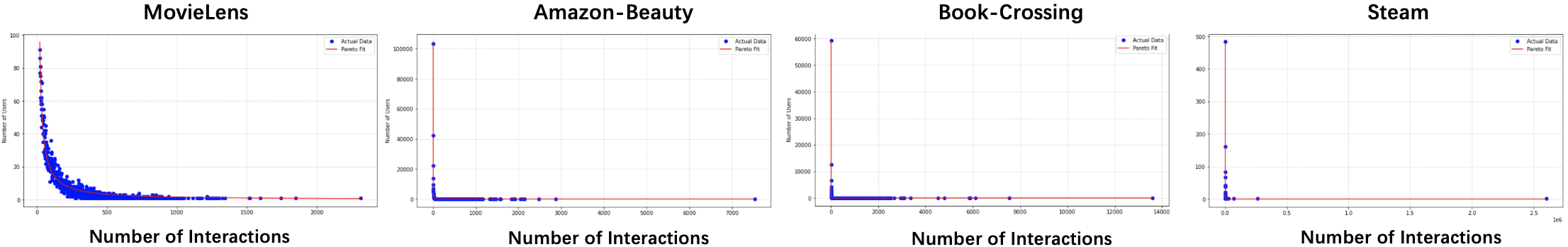}
    }
    \caption{The results of using $p(x)$ to fit real-world datasets. The blue points are the real-world data, and the red lines are the fitted distributions.}
    \label{lt}
\end{figure}

\begin{figure}[t]
    \centering
    \setlength{\fboxrule}{0.pt}
    \setlength{\fboxsep}{0.pt}
    \fbox{
        \includegraphics[width=.8\linewidth]{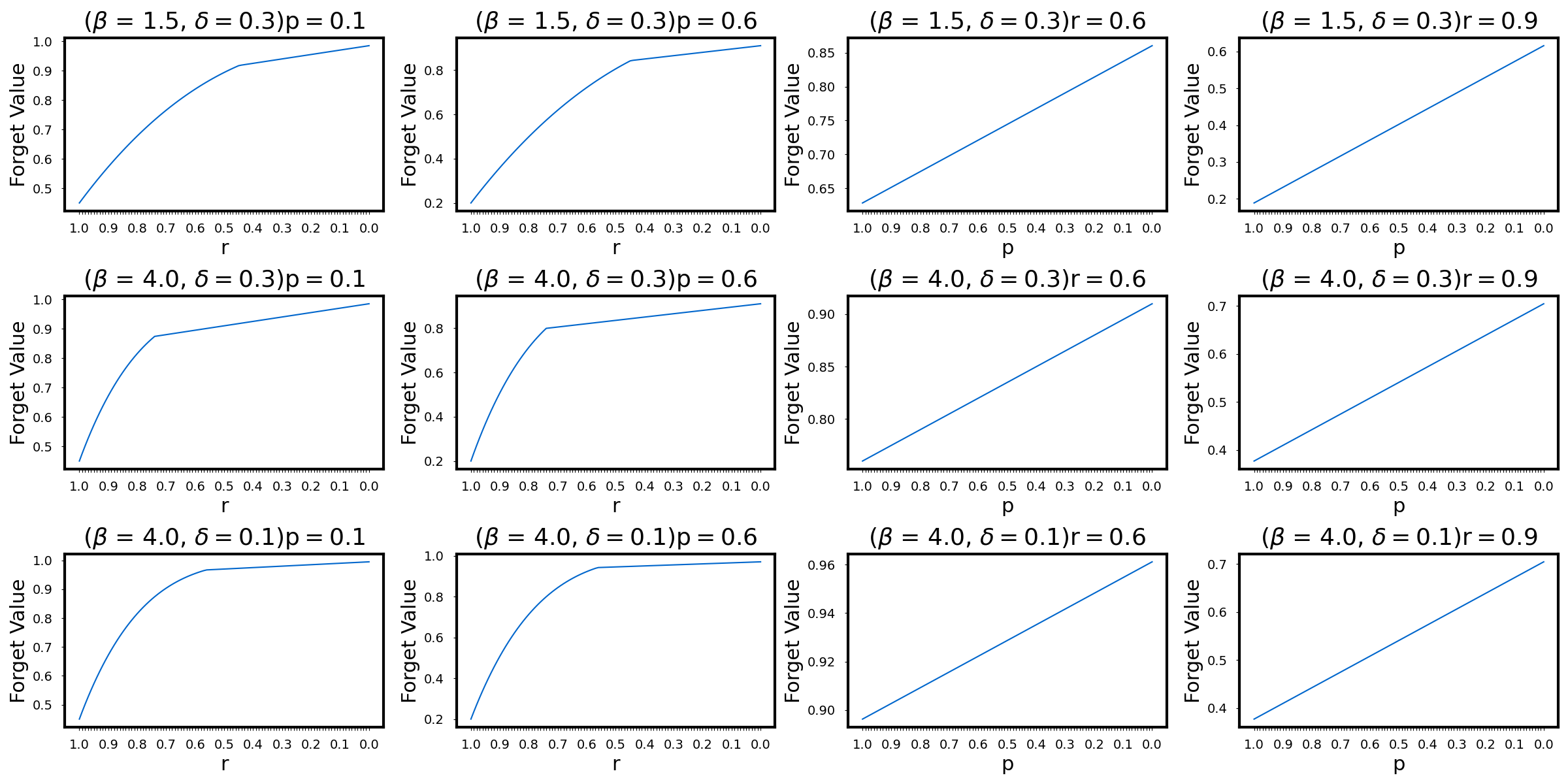}
    }
    \caption{Examples of the curve of $g(M_i)$ with different hyper-parameters.}
    \label{curve}
\end{figure}

\subsection*{Effectiveness of $p(x)$}
To assess whether $p(x)$ can be powerful enough to characterize real-world scenarios, we employ it to fit four well-known recommendation datasets including {MovieLens, Amazon-Beauty, Book-Crossing, and Steam}\footnote{https://github.com/RUCAIBox/RecSysDatasets}, where we approximate the interaction numbers of the users as their activity levels.
The results are illustrated in {Figure~\ref{lt}}. We can see, by varying the values of $\alpha$, $p(x)$ can effectively model the long-tail activity distribution of the real-world datasets.

\subsection*{Example curves of $g(M_i)$}
For an intuitive understanding of the forgotten mechanism in long-term memory, we present several examples of the curve of $g(M_i)$ in Figure~\ref{curve}.

\subsection*{Prompt Examples for different Agent Behaviors}

The prompt used in RecAgent consists of three parts: a summary of the Agent's personal profile, reaction to given observation, and action-specific instructions.

\subsubsection*{Summary}
Summary is the process of extracting and summarizing information relevant to the current observation from the user's profile.
\begin{tcolorbox}[colback=black!3!white,colframe=black!30!white]
    \textbf{\textit{Summary}}:Given the following observation about David Miller: 'David Smith is going to chat with David Miller.', please summarize the relevant details from his profile. His profile information is as follows:

    Name: David Miller

    Age: 39

    Gender:female

    Traits: fun-loving, creative, practical, energetic, patient

    Status: writer

    Movie Interest:  action movies,  sci-fi movies,  classic movies

    Feature: Choose movies, enjoy watching, and provide feedback and ratings to the recommendation system.
    Search for movies heard of before and expand movie experiences{...}Enjoy publicly posting on social media and sharing content and insights with more people.

    Interpersonal Relationships: {'David Smith': 'friend'}

    Please avoid repeating the observation in the summary.

    Summary:
\end{tcolorbox}
\begin{tcolorbox}[colback=black!3!white,colframe=black!30!white]
    \textbf{\textit{Output}}: David Smith is a 25-year-old male photographer who is compassionate, caring, ambitious, and optimistic. He has an interest in sci-fi and comedy movies and enjoys choosing, watching, and providing feedback and ratings to the recommendation system. He has high standards for movies and the recommendation system and may criticize both. He has a friendship with David Miller.
\end{tcolorbox}

\subsection*{Reaction}
This section serves as a shared prompt framework for all actions, integrating information such as summary, memory, observation, and more to obtain the agent's reaction.
It covers reactions for both individual agent actions and dialogues involving two agents.

$\bullet$ Single Agent Reaction
\begin{tcolorbox}[colback=black!3!white,colframe=black!30!white]
    \textbf{\textit{Reaction}}: Name: David Smith (age: 25)
    \{Summary\}

    It is September 12, 2023, 08:00 AM.

    David Smith recently {heard} [\'The Matrix\', \'Back to the Future.\', \'Anchorman\', \'Superbad\'] on social media.

    David Miller recently watched nothing on recommender system.
    Other than that David Miller doesn\'t know any movies.

    Most recent observations: David and Millar had a conversation about their shared interest in movies, discussing their favorite genres such as sci-fi, comedy, and action, exchanging recommendations, expressing their skepticism about recommendation systems, and making plans to compare notes and potentially organize a movie night together.David Miller is seeking recommendations for movies such as "The Matrix," "Back to the Future," "Anchorman," and "Superbad" on social media, expressing interest in planning a movie night and inviting others to share their thoughts and suggestions.

    Observation: David Miller want to post for all acquaintances.

    All occurrences of movie names should be enclosed with <>
\end{tcolorbox}

$\bullet$ Reaction Between Two Agent
\begin{tcolorbox}[colback=black!3!white,colframe=black!30!white]
    \textbf{\textit{Reaction}}:
    Name: David Smith (age: 25)

    {Summary}

    Name: David Miller (age: 39)

    {Summary}

    It is September 12, 2023, 10:28 PM.

    David Smith recently heared [\'Inception\', \'The Dark Knight\', \'Inception\', \'The Dark Knight\'] on social media.

    David Smith recently watched nothing on recommender system.

    Other than that David Smith doesn\'t know any movies.

    David Miller recently {heard} [\'Inception\', \'The Dark Knight\'] on social media.
    David Miller recently watched nothing on recommender system.
    Other than that David Miller doesn\'t know any movies.

    Most recent observations of David Smith: David Smith observed that David Miller was interested in chatting about movies, as evidenced by his action of initiating a chat conversation.David and David had a conversation about their mutual interest in movies, with David Miller recommending sci-fi and action films like "Inception" and "The Dark Knight" to David Smith, who expressed interest in watching them and sharing his thoughts with David Miller in the future.David Miller is excited to watch and review <Inception> and <The Dark Knight> after discovering them on his recommender system.David Smith observed that David Miller had a chat about movie interests, focusing on the reasons behind their preferences.

    Most recent observations of David Miller: David Miller and David Smith had a conversation about their shared interest in movies, with David Miller recommending sci-fi movie "Inception" and action movie "The Dark Knight" to David Smith, who expressed interest in watching them and discussing them later.; David Miller expressed his excitement about two highly recommended movies, "Inception" and "The Dark Knight," which he discovered on social media, and eagerly anticipates watching them and sharing his reviews with others.; David Miller is excited to watch and share his reviews of the movies <Inception> and <The Dark Knight> that he discovered on his recommender system.

    Observation: David Smith is going to chat with David Miller.

    All occurrences of movie names should be enclosed with <>.

\end{tcolorbox}

\subsubsection*{Action}
This section includes all the actions that agents can perform, along with examples of input and output for each action.
$\bullet$ Take Action
\begin{tcolorbox}[colback=black!3!white,colframe=black!30!white]
    \textbf{\textit{OB}}: David Miller must take only ONE of the actions below:
    (1) Enter the Recommender System. If so, David Miller will be recommended some movies, from which David Miller can watch some movies, or search for movies by himself.

    (2) Enter the Social Media. David Miller can chat with friends or publish a post to all friends of David Miller. If David Miller recently watched some movies they might want to share with others.

    (3) Do Nothing.
\end{tcolorbox}
\begin{tcolorbox}[colback=black!3!white,colframe=black!30!white]
    \textbf{\textit{Suffix}}: What action would David Miller like to take? Respond in one line.

    If David Miller wants to enter the Recommender System, write:

    [RECOMMENDER]:: David Miller enters the Recommender System

    If David Miller wants to enter the Social Media, write:

    [SOCIAL]:: David Miller enters the Social Media

    If David Miller wants to do nothing, write:

    [NOTHING]:: David Miller does nothing
\end{tcolorbox}
\begin{tcolorbox}[colback=black!3!white,colframe=black!30!white]
    \textbf{\textit{Output}}: [RECOMMENDER]:: David Miller enters the Recommender System.
\end{tcolorbox}
$\bullet$ Take Recommender Action
\begin{tcolorbox}[colback=black!3!white,colframe=black!30!white]
    \textbf{\textit{OB}}: David Smith is browsing the recommender system. David Smith is recommended ["<Eye of Vichy, The (Oeil de Vichy, L')>||The movie <Eye of Vichy, The (Oeil de Vichy, L') > is a documentary that delves into the propaganda films made by the Vichy government during World War II.", "<Oliver!>||<Oliver!> is a musical adaptation of Charles Dickens' classic novel <Oliver Twist>.", '<...And Justice for All>||<...And Justice for All> is a legal drama film from 1979 that follows the story of an ethical and idealistic defense attorney named Arthur Kirkland, played by Al Pacino. Kirkland finds himself in a moral dilemma when he is forced to defend a judge who he despises, who has been accused of rape. As he navigates the corrupt legal system and grapples with his own values, Kirkland must decide whether to compromise his integrity or risk losing everything he has worked for. The film explores themes of justice, morality, and the flaws of the American legal system.', '<T-Men>||<T-Men> is a film noir about two Treasury agents who go undercover to break up a counterfeiting ring.', '<Son of Flubber>||<Son of Flubber> is a 1963 American comedy film directed by Robert Stevenson and starring Fred MacMurray.'].
\end{tcolorbox}
\begin{tcolorbox}[colback=black!3!white,colframe=black!30!white]
    \textbf{\textit{Suffix}}:
    David Miller must choose one of the four actions below:
    (1) Watch ONLY ONE movie from the list returned by the recommender system.

    (2) See the next page.

    (3) Search for a specific item.

    (4) Leave the recommender system.

    If David Miller has recently heard about a particular movie on social media, David Miller might want to search for that movie on the recommender system.
    To watch a movie from the recommended list, write: [BUY]:: ONLY ONE movie name||description.
    To see the next page, write: [NEXT]:: David Miller views the next page.
    To search for a specific item, write: [SEARCH]:: single, specific movie name to search for.
    To leave the recommender system, write: [LEAVE]:: David Miller leaves the recommender system.

\end{tcolorbox}
\begin{tcolorbox}[colback=black!3!white,colframe=black!30!white]
    \textbf{\textit{Output}}: [BUY]::<Son of Flubber>||<Son of Flubber> is a 1963 American comedy film directed by Robert Stevenson and starring Fred MacMurray.
\end{tcolorbox}
$\bullet$ Generate Feeling
\begin{tcolorbox}[colback=black!3!white,colframe=black!30!white]
    \textbf{\textit{OB}}: David Jones has just finished watching S.F.W.;;S.F.W. (1994) is a dark comedy-drama film about a group of teenagers who are held hostage in a convenience store by a media-savvy, charismatic, and nihilistic young man, Cliff Spab (played by Stephen Dorff), and his equally disillusioned girlfriend, Wendy (played by Reese Witherspoon). The film explores themes of media manipulation, youth disillusionment, and the commodification of tragedy in modern society.
\end{tcolorbox}
\begin{tcolorbox}[colback=black!3!white,colframe=black!30!white]
    \textbf{\textit{Suffix}}: David Jones, how did you feel about the movie you just watched? Describe your feelings in one line.NOTE: Please answer in the first-person perspective.
\end{tcolorbox}
\begin{tcolorbox}[colback=black!3!white,colframe=black!30!white]
    \textbf{\textit{Output}}: I found "SFW" to be a thought-provoking and captivating exploration of media manipulation and youth disillusionment in modern society.
\end{tcolorbox}
$\bullet$ Search Item
\begin{tcolorbox}[colback=black!3!white,colframe=black!30!white]
    \textbf{\textit{OB}}: David Miller is browsing the recommender system. David Miller has searched for Interstellar in recommender system and recommender system returns item list:['Supernova (2000)', 'October Sky (1999)', 'Cat from Outer Space, The (1978)', 'Space Cowboys (2000)', 'Aliens (1986)'] as search results.
\end{tcolorbox}
\begin{tcolorbox}[colback=black!3!white,colframe=black!30!white]
    \textbf{\textit{Suffix}}: David Miller must choose one of the four actions below:

    (1) Watch ONLY ONE movie from the list returned by the recommender system.

    (2) See the next page.

    (3) Search for a specific item.

    (4) Leave the recommender system.

    If David Miller has recently heard about a particular movie on social media, David Miller might want to search for that movie on the recommender system.

    To watch a movie from the recommended list, write:

    [BUY]:: ONLY ONE movie name;;description

    To see the next page, write:

    [NEXT]:: David Miller views the next page.

    To search for a specific item, write:

    [SEARCH]:: single, specific movie name to search for.

    To leave the recommender system, write:

    [LEAVE]:: David Miller leaves the recommender system.
\end{tcolorbox}
\begin{tcolorbox}[colback=black!3!white,colframe=black!30!white]
    \textbf{\textit{Output}}: [SEARCH]:: Inception
\end{tcolorbox}

$\bullet$ Generate Dialogue
\begin{tcolorbox}[colback=black!3!white,colframe=black!30!white]
    \textbf{\textit{OB}}:
    David Smith is chatting with David Miller.
\end{tcolorbox}
\begin{tcolorbox}[colback=black!3!white,colframe=black!30!white]
    \textbf{\textit{Suffix}}:
    What will be said between David Smith and David Miller? David Smith initiates the conversation first. Please simulate their conversation.

    David Smith and David Miller should not say anything about movies they have not watched or heard about.

    Write the dialogue in the following format:

    [David Smith]:

    [David Miller]:

\end{tcolorbox}
\begin{tcolorbox}[colback=black!3!white,colframe=black!30!white]
    \textbf{\textit{Output}}:
    [David Smith]: Hey David! How's it going? I heard you were interested in a movie. What's been on your mind?

    [David Miller]: Hey David! I'm doing great, thanks for asking. Yeah, I've been hearing a lot about this movie <Interstellar> recently. Have you heard of it too?

    [David Smith]: Absolutely! Actually, I've been seeing it all over social media as well. It seems to be a popular choice among movie lovers. I even watched it recently on the recommender system. The storyline was visually stunning and thought-provoking.

    [David Miller]: That's awesome! I'm glad you enjoyed it. I've been wanting to watch it too. Would you be up for a cozy movie night to watch it together? We can discuss our thoughts and interpretations afterwards.

        [David Smith]: I'd love that! It's always more fun to watch movies with friends and have those deep conversations afterwards. Count me in!

    [David Miller]: Great! I'll make sure to set up a movie night soon. By the way, have you heard of any other movies that you'd recommend? I'm always open to expanding my movie experiences.
\end{tcolorbox}

$\bullet$ Publish Posting
\begin{tcolorbox}[colback=black!3!white,colframe=black!30!white]
    \textbf{\textit{OB}}: David Smith want to post for all acquaintances.
\end{tcolorbox}
\begin{tcolorbox}[colback=black!3!white,colframe=black!30!white]
    \textbf{\textit{Suffix}}: Posts should be related to recent watched movies on recommender systems.David Smith should not say anything about movies that have not watched or heard about.

    If you were David Smith, what will you post? Respond in one line.
\end{tcolorbox}
\begin{tcolorbox}[colback=black!3!white,colframe=black!30!white]
    \textbf{\textit{Output}}: Hey everyone! Just watched <Inception> on the recommender system and it was absolutely mind-blowing! Highly recommend checking it out!
\end{tcolorbox}

\bibliographystyle{unsrt}
\bibliography{references}






\clearpage
\setcounter{figure}{0}
\renewcommand{\figurename}{Extended Data Figure}
\setcounter{table}{0}
\renewcommand{\tablename}{Extended Data Table}


\end{document}